\documentstyle[twocolumn,aps,prl]{revtex}
\begin{document}
\draft
\title{Spontaneous Chiral Symmetry Breaking Beyond BCS}
\author{Pedro J. de A. Bicudo} 
\address{Departamento de F\'isica and Centro de F\'isica das 
Interac\c c\~oes Fundamentais, Edif\'icio Ci\^encia, Instituto 
Superior T\'ecnico, Av. Rovisco Pais, 
1096 Lisboa, Portugal}
\maketitle
\begin{abstract}  
At the BCS level of chiral symmetry breaking, the mass gap equation for
quark-antiquark
condensation only uses the kernel of the Bethe-Salpeter equation.
We introduce coupled channels with ladder mesons in the
mass gap equation. Consistency is insured by the Ward Identity
for axial currents, and the $\pi$ remains a Goldstone boson in the 
chiral limit. We find that bare mesons with confined quarks do not 
contribute to the mass gap equation, and estimate that the contribution
of full mesons may be large. 
\end{abstract}
\pacs{11.30.Rd, 74.20.Fg, 24.10.Eq, 14.40.Aq}
%
%
\par
The studying of dynamical symmetry breaking beyond BCS has never been 
done before. 
This amounts to join the 
BCS mechanism with the mean field expansion of effective bound states. 
Whenever a field theory produces bound state solutions, one
should try to include them at the onset in the self-consistent Schwinger 
Dyson equations of that theory.
The solution of this technical problem is not only relevant
from condensed matter physics to particle physics,
but it is essential for a correct understanding of hadronic physics.
\par
Strong interactions were observed, at the very beginning of hadronic 
physics, with large scattering lengths and wide decay widths. 
The discovery of quarks as the microscopic elements of 
hadrons, and two new theoretical concepts, confinement and chiral 
symmetry breaking, eclipsed partially the importance of coupled 
channels in theoretical hadronic physics.
Several different strong or confining, chiral invariant quark-antiquark 
interactions have been used
to condense in the vacuum scalar quark-antiquark pairs, 
inducing dynamical chiral symmetry breaking, 
and a Goldstone boson, the $\pi$ was found \cite{Nambu}, 
in the case of vanishing quark masses.
Recently coupled channel effects have been reevaluated in 
many hadronic phenomena. 
A paradigmatic case is the omnipresent nucleon, where the 
estimated negative 
mass shift due to the coupling to  the channels 
$\pi \ N$, $\pi \ \Delta$, $\pi \ N* \dots$  
has increased almost an order of magnitude in the last decade
 \cite{Thomas,Richard}.
Moreover in the literature only 1 or 2 coupled channels are usually 
included, and it remains to be proved theoretically that the
total mass shift due to the infinite series of coupled channels is finite.
\par
The crucial problem seems to dwell in the pion mass \cite{Emilio}. 
The pion itself is coupled to the channels 
$\pi \ \rho, \ \rho \ \rho \cdots $ 
and his bare mass is believed to suffer a large shift of the $GeV$ order.
This effect is so large \cite{Tornqvist}, 
 that it disputes with the instanton \cite{Hooft} the
solution of the $U(1)$ problem of the $\pi \ \eta$ mass shift. Now
suppose that the mass gap equation for chiral symmetry breaking was
solved at the BCS level, i.e. without including the coupled channels,
then a bare pion with vanishing bare mass would be a solution of the
Bethe Salpeter equation. If the coupled 
channels were then included, at posteriori, in the bound state equation then
the pion mass would be the sum of the small bare mass plus a large negative
mass shift and thus would have a resulting negative mass, which is absurd.
We will have to use \cite{Schrieffer} the Ward identities (WI) in order to 
insure that the 
bound state equation for the pion -a Bethe Salpeter equation with coupled 
channels - is consistent with the non linear mass gap equation. A reward
of this study is a $\pi$ with a vanishing positive mass in the chiral limit, 
and a tower of states above the $\pi$ with higher masses \cite{Emilio}.
%
%
\par
The WI were first derived for QED, and concerned 
the free vector vertex $\Gamma^{\mu}_f= \gamma^{\mu}$, 
and the free Dirac fermion 
propagator $\displaystyle {\cal S}_f(k)= i/ \not{k}-m+i\epsilon$. 
There is also a WI identity for the free axial vector vertex 
$\Gamma^{\mu 5}_f= \gamma^{\mu}\gamma^5$ that involves the free
pseudoscalar vertex $\Gamma^5_f= \gamma^5$,
\FL
\begin{eqnarray}
\label{axial WI}
&&-i (k_{\mu}-k_{\mu}') {\cal S}(k)\Gamma^{\mu 5}(k-k'){\cal S}(k')
\nonumber \\ &&+
2im{\cal S}(k)\Gamma^5(k-k'){\cal S}(k') 
= {\cal S}(k)\gamma^5 + \gamma^5{\cal S}(k'),    
\end{eqnarray}
where the difference in the right member of the equation 
extends the identity to renormalized propagators 
and vertices. 
The WI are crucial to ensure the consistency 
between the mass gap equation (MGE) and Bethe-Salpeter (BS) equation. 
They ensure that 
the self energy of the MGE is obtained from the BS kernel 
by closing the fermion line where
the vertex is inserted, and that no double counting occurs. 
Inversely they also ensure
that the BS kernel is obtained if one inserts the vertex in all possible
propagators of the self energy. 
\par
This will now be illustrated at the BCS level.
In this case the mass gap equation and the BS equation are, 
\begin{eqnarray}  
\label{BCS MGE} \displaystyle
-\hspace{-.15cm}-\hspace{-.2cm}\longleftarrow  ^{-1} 
\ &=& \ {\cal S}_0^{-1} \    
- \ { ^{\cdot \hspace{-.02cm} ^{\cdot^{\cdot}}}} 
{^{^{ ^{\cdot}}}} {^{^{\cdot}}} \hspace{-.02cm} {^{\cdot} }
\hspace{-.6cm}-\hspace{-.2cm}\leftarrow   
\\ \label{BCS BSE}
\bullet \hspace{-.2cm} = \ \ \ \  &=& \ \Gamma_0 \ \ + \
{{\atop{\cdot \atop \cdot}} \atop { {\cdot \atop \cdot} \atop } }
\hspace{-.15cm} {\nwarrow \atop \nearrow} \hspace{-.1cm}
\bullet \hspace{-.2cm} =
\end{eqnarray}
where the full propagator is 
denoted  by $ -\hspace{-.1cm}-\hspace{-.2cm}\longleftarrow $ 
and the vertex will be denoted 
by $ \bullet \hspace{-.2cm} = \hspace{-.1cm} =$.  
The effective quark-quark 2-body interaction, a chiral invariant and 
color dependent interaction, is represented with a 
dotted line $ \dots \dots$. 
With the WI, it is possible to check directly that these eqs. are 
equivalent, both for dressed and for free propagators and vertices. 
Nevertheless we will now illustrate explicitly how the free WI recovers 
the MGE from the BS equation. 
Up to second order in the potential insertions, 
the vertex obtained from eq. (\ref{BCS BSE}) is,  
\FL
\begin{eqnarray}
\label{2OV}
\begin{picture}(12,10)(0,3)
\put(0,0){$_{k'}$}
\put(0,10){$_{k}$}
\end{picture}
\begin{picture}(10,2)(0,-1)
\put(-5,-1.5){$\bullet$}
\put(0,0){\line(1,0){10}}
\put(0,2){\line(1,0){10}}
\end{picture}
\
&=& \Gamma_0  +  \
\begin{picture}(10,10)(0,0)
\put(0,0){$_{k'}$}
\put(0,10){$_{k}$}
\end{picture}
\begin{picture}(30,10)(0,3)
\put(0,0){\vector(1,0){10}}
\put(10,0){\line(1,0){5}}
\put(7,-5){$_0$}
\put(15,10){\vector(-1,0){10}}
\put(5,10){\line(-1,0){5}}
\put(7,5){$_0$}
\multiput(0,-2)(0,2){6}{$\cdot$}
\put(15,3){\oval(6,6)[br]}
\put(15,7){\oval(6,6)[tr]}
\put(20,2){$\Gamma_0$}
\end{picture}
\ + \
\begin{picture}(10,20)(0,0)
\put(0,0){$_{k'}$}
\put(0,10){$_{k}$}
\end{picture}
\begin{picture}(60,20)(0,3)
\put(15,10){\vector(-1,0){10}}
\put(5,10){\line(-1,0){5}}
\put(7,5){$_0$}
\multiput(0,-2)(0,2){6}{$\cdot$}
\put(0,0){\vector(1,0){25}}
\put(25,0){\line(1,0){25}}
\put(25,-5){$_0$}
\put(15,10){\begin{picture}(20,10)(0,0)
\put(0,0){\line(1,0){10}}
\put(20,0){\vector(-1,0){10}}
\put(10,-5){$_0$}
\put(0,0){$\cdot$}
\put(1,2.2){$\cdot$}
\put(3,3.5){$\cdot$}
\put(5.6,4.5){$\cdot$}
\put(10,5){$\cdot$}
\put(14.4,4.5){$\cdot$}
\put(17,3.5){$\cdot$}
\put(19,2.2){$\cdot$}
\put(20,0){$\cdot$}
\end{picture}}
\put(25,5){$_0$}
\put(50,10){\vector(-1,0){10}}
\put(40,10){\line(-1,0){5}}
\put(40,5){$_0$}
\put(50,3){\oval(6,6)[br]}
\put(50,7){\oval(6,6)[tr]}
\put(55,2){$\Gamma_0$}
\end{picture}
\nonumber \\ 
&&+ \
\begin{picture}(10,10)(0,0)
\put(0,0){$_{k'}$}
\put(0,10){$_{k}$}
\end{picture}
\begin{picture}(40,10)(0,3)
\put(0,0){\vector(1,0){10}}
\put(10,0){\line(1,0){5}}
\put(7,-5){$_0$}
\put(15,10){\vector(-1,0){10}}
\put(5,10){\line(-1,0){5}}
\put(7,5){$_0$}
\multiput(0,-2)(0,2){6}{$\cdot$}
\put(15,0){\vector(1,0){10}}
\put(25,0){\line(1,0){5}}
\put(22,-5){$_0$}
\put(30,10){\vector(-1,0){10}}
\put(20,10){\line(-1,0){5}}
\put(22,5){$_0$}
\multiput(15,-2)(0,2){6}{$\cdot$}
\put(30,3){\oval(6,6)[br]}
\put(30,7){\oval(6,6)[tr]}
\put(35,2){$\Gamma_0$}
\end{picture}
\ + \
\begin{picture}(10,20)(0,0)
\put(0,0){$_{k'}$}
\put(0,10){$_{k}$}
\end{picture}
\begin{picture}(60,20)(0,3)
\put(0,0){\vector(1,0){10}}
\put(10,0){\line(1,0){5}}
\put(7,-5){$_0$}
\multiput(0,-2)(0,2){6}{$\cdot$}
\put(50,10){\vector(-1,0){25}}
\put(25,10){\line(-1,0){25}}
\put(25,15){$_0$}
\put(15,0){\begin{picture}(20,10)(0,0)
\put(0,0){\vector(1,0){10}}
\put(20,0){\line(-1,0){10}}
\put(10,-5){$_0$}
\put(0,0){$\cdot$}
\put(1,2.2){$\cdot$}
\put(3,3.5){$\cdot$}
\put(5.6,4.5){$\cdot$}
\put(10,5){$\cdot$}
\put(14.4,4.5){$\cdot$}
\put(17,3.5){$\cdot$}
\put(19,2.2){$\cdot$}
\put(20,0){$\cdot$}
\end{picture}}
\put(35,0){\vector(1,0){10}}
\put(45,0){\line(1,0){5}}
\put(40,-5){$_0$}
\put(50,3){\oval(6,6)[br]}
\put(50,7){\oval(6,6)[tr]}
\put(55,2){$\Gamma_0$}
\end{picture}
\ + \dots 
\end{eqnarray}
Substituting the WI in eq(\ref{2OV}), we get,
\FL
\begin{eqnarray}
&&\gamma_5{\cal S}^{-1}(k') + {\cal S}^{-1}(k)\gamma_5= 
\gamma_5{\cal S}^{-1}_0(k') + {\cal S}^{-1}_0(k)\gamma_5 
\ + 
\begin{picture}(20,10)(0,0)
\put(0,0){\line(1,0){10}}
\put(20,-8){$_0$}
\put(20,0){\vector(-1,0){10}}
\put(10,-5){$_{k}$}
\put(17,-2.5){$\ast$}
\put(0,0){$\cdot$}
\put(1,4.4){$\cdot$}
\put(3,7){$\cdot$}
\put(5.6,9){$\cdot$}
\put(10,10){$\cdot$}
\put(14.4,9){$\cdot$}
\put(17,7){$\cdot$}
\put(19,4.4){$\cdot$}
\put(20,0){$\cdot$}
\end{picture}  
\nonumber  \\ && + \
\begin{picture}(20,10)(0,0)
\put(20,-8){$_0$}
\put(0,0){\line(1,0){10}}
\put(20,0){\vector(-1,0){10}}
\put(12,-5){$_{k'}$}
\put(1,-2.5){$\ast$}
\put(0,0){$\cdot$}
\put(1,4.4){$\cdot$}
\put(3,7){$\cdot$}
\put(5.6,9){$\cdot$}
\put(10,10){$\cdot$}
\put(14.4,9){$\cdot$}
\put(17,7){$\cdot$}
\put(19,4.4){$\cdot$}
\put(20,0){$\cdot$}
\end{picture}  
\ + \
\begin{picture}(50,25)(0,0)
\put(50,-8){$_0$}
\put(15,0){\vector(-1,0){10}}
\put(5,0){\line(-1,0){5}}
\put(7,-5){$_{k}$}
\put(50,0){\vector(-1,0){10}}
\put(40,0){\line(-1,0){5}}
\put(42,-5){$_{k}$}
\put(47,-2.5){$\ast$}
\put(50.00,0.00){$\cdot$}
\put(49.69,1.96){$\cdot$}
\put(48.77,3.86){$\cdot$}
\put(47.27,5.92){$\cdot$}
\put(45.22,7.20){$\cdot$}
\put(42.67,8.84){$\cdot$}
\put(39.69,10.11){$\cdot$}
\put(36.34,11.14){$\cdot$}
\put(32.72,11.89){$\cdot$}
\put(28.91,12.35){$\cdot$}
\put(25.00,12.50){$\cdot$}
\put(21.08,12.35){$\cdot$}
\put(17.27,11.89){$\cdot$}
\put(13.65,11.14){$\cdot$}
\put(10.30,10.11){$\cdot$}
\put(7.32,8.84){$\cdot$}
\put(4.77,7.20){$\cdot$}
\put(2.72,5.92){$\cdot$}
\put(1.22,3.86){$\cdot$}
\put(0.30,1.96){$\cdot$}
\put(0.00,0.00){$\cdot$}
\put(15,0){\begin{picture}(20,10)(0,0)
\put(0,0){\line(1,0){10}}
\put(20,0){\vector(-1,0){10}}
\put(10,-5){$_{k}$}
\put(0,0){$\cdot$}
\put(1,2.2){$\cdot$}
\put(3,3.5){$\cdot$}
\put(5.6,4.5){$\cdot$}
\put(10,5){$\cdot$}
\put(14.4,4.5){$\cdot$}
\put(17,3.5){$\cdot$}
\put(19,2.2){$\cdot$}
\put(20,0){$\cdot$}
\end{picture}}
\end{picture}  
\ + \
\begin{picture}(50,25)(0,0)
\put(50,-8){$_0$}
\put(15,0){\vector(-1,0){10}}
\put(5,0){\line(-1,0){5}}
\put(7,-5){$_{k}$}
\put(50,0){\vector(-1,0){10}}
\put(40,0){\line(-1,0){5}}
\put(42,-5){$_{k'}$}
\put(36,-2.5){$\ast$}
\put(50.00,0.00){$\cdot$}
\put(49.69,1.96){$\cdot$}
\put(48.77,3.86){$\cdot$}
\put(47.27,5.92){$\cdot$}
\put(45.22,7.20){$\cdot$}
\put(42.67,8.84){$\cdot$}
\put(39.69,10.11){$\cdot$}
\put(36.34,11.14){$\cdot$}
\put(32.72,11.89){$\cdot$}
\put(28.91,12.35){$\cdot$}
\put(25.00,12.50){$\cdot$}
\put(21.08,12.35){$\cdot$}
\put(17.27,11.89){$\cdot$}
\put(13.65,11.14){$\cdot$}
\put(10.30,10.11){$\cdot$}
\put(7.32,8.84){$\cdot$}
\put(4.77,7.20){$\cdot$}
\put(2.72,5.92){$\cdot$}
\put(1.22,3.86){$\cdot$}
\put(0.30,1.96){$\cdot$}
\put(0.00,0.00){$\cdot$}
\put(15,0){\begin{picture}(20,10)(0,0)
\put(0,0){\line(1,0){10}}
\put(20,0){\vector(-1,0){10}}
\put(10,-5){$_{k}$}
\put(0,0){$\cdot$}
\put(1,2.2){$\cdot$}
\put(3,3.5){$\cdot$}
\put(5.6,4.5){$\cdot$}
\put(10,5){$\cdot$}
\put(14.4,4.5){$\cdot$}
\put(17,3.5){$\cdot$}
\put(19,2.2){$\cdot$}
\put(20,0){$\cdot$}
\end{picture}}
\end{picture}  
\ + \
\begin{picture}(50,25)(0,0)
\put(50,-8){$_0$}
\put(15,0){\vector(-1,0){10}}
\put(5,0){\line(-1,0){5}}
\put(7,-5){$_{k}$}
\put(50,0){\vector(-1,0){10}}
\put(40,0){\line(-1,0){5}}
\put(42,-5){$_{k'}$}
\put(32,-2.5){$\ast$}
\put(50.00,0.00){$\cdot$}
\put(49.69,1.96){$\cdot$}
\put(48.77,3.86){$\cdot$}
\put(47.27,5.92){$\cdot$}
\put(45.22,7.20){$\cdot$}
\put(42.67,8.84){$\cdot$}
\put(39.69,10.11){$\cdot$}
\put(36.34,11.14){$\cdot$}
\put(32.72,11.89){$\cdot$}
\put(28.91,12.35){$\cdot$}
\put(25.00,12.50){$\cdot$}
\put(21.08,12.35){$\cdot$}
\put(17.27,11.89){$\cdot$}
\put(13.65,11.14){$\cdot$}
\put(10.30,10.11){$\cdot$}
\put(7.32,8.84){$\cdot$}
\put(4.77,7.20){$\cdot$}
\put(2.72,5.92){$\cdot$}
\put(1.22,3.86){$\cdot$}
\put(0.30,1.96){$\cdot$}
\put(0.00,0.00){$\cdot$}
\put(15,0){\begin{picture}(20,10)(0,0)
\put(0,0){\line(1,0){10}}
\put(20,0){\vector(-1,0){10}}
\put(10,-5){$_{k}$}
\put(0,0){$\cdot$}
\put(1,2.2){$\cdot$}
\put(3,3.5){$\cdot$}
\put(5.6,4.5){$\cdot$}
\put(10,5){$\cdot$}
\put(14.4,4.5){$\cdot$}
\put(17,3.5){$\cdot$}
\put(19,2.2){$\cdot$}
\put(20,0){$\cdot$}
\end{picture}}
\end{picture}  
\nonumber \\ && + \
\begin{picture}(50,25)(0,0)
\put(50,-8){$_0$}
\put(15,0){\vector(-1,0){10}}
\put(5,0){\line(-1,0){5}}
\put(7,-5){$_{k}$}
\put(50,0){\vector(-1,0){10}}
\put(40,0){\line(-1,0){5}}
\put(42,-5){$_{k'}$}
\put(16,-2.5){$\ast$}
\put(50.00,0.00){$\cdot$}
\put(49.69,1.96){$\cdot$}
\put(48.77,3.86){$\cdot$}
\put(47.27,5.92){$\cdot$}
\put(45.22,7.20){$\cdot$}
\put(42.67,8.84){$\cdot$}
\put(39.69,10.11){$\cdot$}
\put(36.34,11.14){$\cdot$}
\put(32.72,11.89){$\cdot$}
\put(28.91,12.35){$\cdot$}
\put(25.00,12.50){$\cdot$}
\put(21.08,12.35){$\cdot$}
\put(17.27,11.89){$\cdot$}
\put(13.65,11.14){$\cdot$}
\put(10.30,10.11){$\cdot$}
\put(7.32,8.84){$\cdot$}
\put(4.77,7.20){$\cdot$}
\put(2.72,5.92){$\cdot$}
\put(1.22,3.86){$\cdot$}
\put(0.30,1.96){$\cdot$}
\put(0.00,0.00){$\cdot$}
\put(15,0){\begin{picture}(20,10)(0,0)
\put(0,0){\line(1,0){10}}
\put(20,0){\vector(-1,0){10}}
\put(10,-5){$_{k'}$}
\put(0,0){$\cdot$}
\put(1,2.2){$\cdot$}
\put(3,3.5){$\cdot$}
\put(5.6,4.5){$\cdot$}
\put(10,5){$\cdot$}
\put(14.4,4.5){$\cdot$}
\put(17,3.5){$\cdot$}
\put(19,2.2){$\cdot$}
\put(20,0){$\cdot$}
\end{picture}}
\end{picture}  
\ + \
\begin{picture}(50,25)(0,0)
\put(50,-8){$_0$}
\put(15,0){\vector(-1,0){10}}
\put(5,0){\line(-1,0){5}}
\put(7,-5){$_{k}$}
\put(50,0){\vector(-1,0){10}}
\put(40,0){\line(-1,0){5}}
\put(42,-5){$_{k'}$}
\put(12,-2.5){$\ast$}
\put(50.00,0.00){$\cdot$}
\put(49.69,1.96){$\cdot$}
\put(48.77,3.86){$\cdot$}
\put(47.27,5.92){$\cdot$}
\put(45.22,7.20){$\cdot$}
\put(42.67,8.84){$\cdot$}
\put(39.69,10.11){$\cdot$}
\put(36.34,11.14){$\cdot$}
\put(32.72,11.89){$\cdot$}
\put(28.91,12.35){$\cdot$}
\put(25.00,12.50){$\cdot$}
\put(21.08,12.35){$\cdot$}
\put(17.27,11.89){$\cdot$}
\put(13.65,11.14){$\cdot$}
\put(10.30,10.11){$\cdot$}
\put(7.32,8.84){$\cdot$}
\put(4.77,7.20){$\cdot$}
\put(2.72,5.92){$\cdot$}
\put(1.22,3.86){$\cdot$}
\put(0.30,1.96){$\cdot$}
\put(0.00,0.00){$\cdot$}
\put(15,0){\begin{picture}(20,10)(0,0)
\put(0,0){\line(1,0){10}}
\put(20,0){\vector(-1,0){10}}
\put(10,-5){$_{k'}$}
\put(0,0){$\cdot$}
\put(1,2.2){$\cdot$}
\put(3,3.5){$\cdot$}
\put(5.6,4.5){$\cdot$}
\put(10,5){$\cdot$}
\put(14.4,4.5){$\cdot$}
\put(17,3.5){$\cdot$}
\put(19,2.2){$\cdot$}
\put(20,0){$\cdot$}
\end{picture}}
\end{picture}  
\ + \
\begin{picture}(50,25)(0,0)
\put(50,-8){$_0$}
\put(15,0){\vector(-1,0){10}}
\put(5,0){\line(-1,0){5}}
\put(7,-5){$_{k'}$}
\put(50,0){\vector(-1,0){10}}
\put(40,0){\line(-1,0){5}}
\put(42,-5){$_{k'}$}
\put(1,-2.5){$\ast$}
\put(50.00,0.00){$\cdot$}
\put(49.69,1.96){$\cdot$}
\put(48.77,3.86){$\cdot$}
\put(47.27,5.92){$\cdot$}
\put(45.22,7.20){$\cdot$}
\put(42.67,8.84){$\cdot$}
\put(39.69,10.11){$\cdot$}
\put(36.34,11.14){$\cdot$}
\put(32.72,11.89){$\cdot$}
\put(28.91,12.35){$\cdot$}
\put(25.00,12.50){$\cdot$}
\put(21.08,12.35){$\cdot$}
\put(17.27,11.89){$\cdot$}
\put(13.65,11.14){$\cdot$}
\put(10.30,10.11){$\cdot$}
\put(7.32,8.84){$\cdot$}
\put(4.77,7.20){$\cdot$}
\put(2.72,5.92){$\cdot$}
\put(1.22,3.86){$\cdot$}
\put(0.30,1.96){$\cdot$}
\put(0.00,0.00){$\cdot$}
\put(15,0){\begin{picture}(20,10)(0,0)
\put(0,0){\line(1,0){10}}
\put(20,0){\vector(-1,0){10}}
\put(10,-5){$_{k'}$}
\put(0,0){$\cdot$}
\put(1,2.2){$\cdot$}
\put(3,3.5){$\cdot$}
\put(5.6,4.5){$\cdot$}
\put(10,5){$\cdot$}
\put(14.4,4.5){$\cdot$}
\put(17,3.5){$\cdot$}
\put(19,2.2){$\cdot$}
\put(20,0){$\cdot$}
\end{picture}}
\end{picture}  
\ + \dots 
\end{eqnarray}
where the Dirac $\gamma_5$ was represented in the diagrams with an $\ast$,
the loop variables were omitted and we only showed 
the $k$ or $k'$ dependence of the propagators.
In the case of a chiral invariant interaction,
the vertices anticommute with $\gamma_5$, 
which cancels exactly the spurious diagrams
depending on both $k$ and $k'$, and we recover the 
propagator of eq. (\ref{BCS MGE})
up to second order in the potential insertions. 
\par
A key product of the WI is the proof that a pseudoscalar  
Goldstone boson exists when current quark masses vanish. 
We will follow a variant of Pagel's proof \cite{Pagels} which 
also yields the BS amplitude itself.
When dynamical symmetry breaking occurs, 
the full propagator is renormalized and the selfconsistent
equation for the self-energy $\Sigma$
has a solution with a finite dynamical quark mass,
\begin{equation}
\label{self}
{\cal S}^{-1}(k)={\cal S}_0^{-1}(k)-\Sigma(k) \ , \ i \Sigma =
 A(k)-\not k B(k)  \ .
\end{equation}
If we substitute this propagator in the WI, we find the solution for
the pseudoscalar vertex $\Gamma^5$ with a vanishing $k-k'$,
\begin{equation}
\Gamma^5(k = k') = { A(k)+m \over m} \gamma^5 
\end{equation}
which diverges for a vanishing quark mass $m$ and shows 
that the pole of a massless pseudoscalar meson appears in the
axial vector vertex, with a boundstate amplitude 
$ A(p) \gamma^5 \over f_\pi$ where $f_\pi$ is a norm.
Because of the axial anomaly, the identity (\ref{axial WI}) for flavour
singlet currents is not verified in this form, and it is necessary
to extend (trivially) this relation to flavoured currents 
in order to keep track of the Goldstone boson. 
%
%
\par
The first step to go beyond BCS is to include a ladder exchange
potential in the self energy $\Sigma$, which ladder includes a
series of meson exchange interactions.
We also include an extra effective interaction, which simplifies the
the energy dependence of $\Sigma$, and limits the meson couplings
in the bound state interactions to vertices of 3 mesons.
The mass gap equation is now,
\FL
\begin{equation}
\label{CS3 self}
\Sigma  =  
\begin{picture}(20,10)(0,0)
\put(0,0){\line(1,0){10}}
\put(20,0){\vector(-1,0){10}}
\put(0,0){$\cdot$}
\put(1,4.4){$\cdot$}
\put(3,7){$\cdot$}
\put(5.6,9){$\cdot$}
\put(10,10){$\cdot$}
\put(14.4,9){$\cdot$}
\put(17,7){$\cdot$}
\put(19,4.4){$\cdot$}
\put(20,0){$\cdot$}
\end{picture}  
\ + 
\begin{picture}(50,25)(0,0)
\put(15,0){\vector(-1,0){10}}
\put(5,0){\line(-1,0){5}}
\put(50,0){\vector(-1,0){10}}
\put(40,0){\line(-1,0){5}}
\put(50.00,0.00){$\cdot$}
\put(49.69,1.96){$\cdot$}
\put(48.77,3.86){$\cdot$}
\put(47.27,5.92){$\cdot$}
\put(45.22,7.20){$\cdot$}
\put(42.67,8.84){$\cdot$}
\put(39.69,10.11){$\cdot$}
\put(36.34,11.14){$\cdot$}
\put(32.72,11.89){$\cdot$}
\put(28.91,12.35){$\cdot$}
\put(25.00,12.50){$\cdot$}
\put(21.08,12.35){$\cdot$}
\put(17.27,11.89){$\cdot$}
\put(13.65,11.14){$\cdot$}
\put(10.30,10.11){$\cdot$}
\put(7.32,8.84){$\cdot$}
\put(4.77,7.20){$\cdot$}
\put(2.72,5.92){$\cdot$}
\put(1.22,3.86){$\cdot$}
\put(0.30,1.96){$\cdot$}
\put(0.00,0.00){$\cdot$}
\put(18,-2){$\Sigma_s$}
\end{picture}  
\ , \
\Sigma_s =
\begin{picture}(75,25)(0,0)
\put(15,0){\line(1,0){15}}
\put(40,0){\vector(-1,0){15}}
\put(15,5){\oval(10,10)[l]}
\put(15,5){\oval(30,30)[lt]}
\put(40,5){\oval(30,10)[r]}
\put(40,5){\oval(50,30)[rt]}
\multiput(-2,3)(2,0){6}{$\cdot$}
\put(15,20){\line(1,0){5}}
\put(15,10){\vector(1,0){10}}
\put(30,20){\vector(-1,0){10}}
\put(30,10){\line(-1,0){5}}
\put(30,5){\framebox(10,20){}}
\put(50,20){\vector(-1,0){5}}
\put(65,0){${\cal S}^{-1}$}
\end{picture},
\end{equation}
where 
$\begin{picture}(40,8)
\put(0,7){\line(1,0){5}}
\put(0,1){\vector(1,0){10}}
\put(15,7){\vector(-1,0){10}}
\put(15,1){\line(-1,0){5}}
\put(15,-1){\framebox(10,10){}}
\put(25,7){\line(1,0){5}}
\put(25,1){\vector(1,0){10}}
\put(40,7){\vector(-1,0){10}}
\put(40,1){\line(-1,0){5}}
\end{picture}$
represents a ladder of effective potential insertions in a pair of
quark and antiquark lines.
Using the prescription of inserting the vertex in all possible
propagators we arrive at the corresponding bound state equation for the
vertex,
\FL
\begin{eqnarray}
\label{CS3 kernel}
\begin{picture}(17,10)(0,0)
\put(2,0){
\begin{picture}(10,2)(0,-1)
\put(-5,-1.5){$\bullet$}
\put(0,0){\line(1,0){10}}
\put(0,2){\line(1,0){10}}
\end{picture}
}\end{picture}
\ &=& \
\Gamma_0  \ + \
\begin{picture}(30,10)(0,0)
\put(0,0){\vector(1,0){10}}
\put(10,0){\line(1,0){5}}
\put(15,5){\oval(10,10)[r]}
\put(15,10){\vector(-1,0){10}}
\put(5,10){\line(-1,0){5}}
\multiput(0,-2)(0,2){6}{$\cdot$}
\put(20,5){
\begin{picture}(10,2)(0,1)
\put(-5,-1.5){$\bullet$}
\put(0,0){\line(1,0){10}}
\put(0,2){\line(1,0){10}}
\end{picture}}
\end{picture}
\ +
\nonumber \\ &&
\begin{picture}(50,15)(0,0)
\put(0,15){
\begin{picture}(30,20)(0,0)
\multiput(0,3)(0,2){6}{$\cdot$}
\put(0,5){\vector(1,0){10}}
\put(15,5){\line(-1,0){5}}
\put(0,15){\line(1,0){5}}
\put(15,15){\vector(-1,0){10}}
\end{picture}}
\put(15,15){
\begin{picture}(30,20)(0,0)
\put(15,10){\oval(10,10)[r]}
\put(0,12){$\Sigma_s$}
\put(15,15){\line(-1,0){7}}
\put(15,15){\vector(-1,0){5}}
\put(0,5){\line(1,0){15}}
\end{picture}}
\put(38,25){
\begin{picture}(10,2)(0,1)
\put(-5,-1.5){$\bullet$}
\put(0,0){\line(1,0){10}}
\put(0,2){\line(1,0){10}}
\end{picture}}
\end{picture}
+
\begin{picture}(50,15)(0,0)
\put(0,15){
\begin{picture}(30,20)(0,0)
\multiput(0,3)(0,2){6}{$\cdot$}
\put(0,5){\vector(1,0){10}}
\put(15,5){\line(-1,0){5}}
\put(0,15){\line(1,0){5}}
\put(15,15){\vector(-1,0){10}}
\end{picture}}
\put(15,15){
\begin{picture}(30,20)(0,0)
\put(15,10){\oval(10,10)[r]}
\put(0,2){$\Sigma_s$}
\put(0,15){\line(1,0){15}}
\put(8,5){\vector(1,0){5}}
\put(8,5){\line(1,0){7}}
\end{picture}}
\put(38,25){
\begin{picture}(10,2)(0,1)
\put(-5,-1.5){$\bullet$}
\put(0,0){\line(1,0){10}}
\put(0,2){\line(1,0){10}}
\end{picture}}
\end{picture}
+
\begin{picture}(80,50)(0,0)
\put(0,15){
\begin{picture}(30,20)(0,0)
\multiput(0,3)(0,2){6}{$\cdot$}
\put(0,5){\vector(1,0){10}}
\put(15,5){\line(-1,0){5}}
\put(0,15){\line(1,0){5}}
\put(15,15){\vector(-1,0){10}}
\end{picture}}
\put(15,0){
\begin{picture}(15,50)(0,0)
\put(15,35){\oval(20,20)[tl]}
\put(15,15){\oval(20,20)[bl]}
\put(0,35){\oval(10,10)[br]}
\put(0,15){\oval(10,10)[tr]}
\put(15,25){\oval(10,20)[l]}
\put(10,20){\vector(0,1){5}}
\end{picture} }
\put(55,0){
\begin{picture}(10,50)(0,0)
\put(0,25){\oval(10,20)[r]}
\put(0,25){\oval(20,40)[r]}
\put(10,15){\vector(0,1){5}}
\put(10,30){\vector(0,1){5}}
\put(5,25){\vector(0,-1){5}}
\end{picture}}
\put(30,0){
\begin{picture}(25,20)(0,0)
\put(0,0){\framebox(10,20){}}
\put(10,5){\vector(1,0){10}}
\put(25,5){\line(-1,0){5}}
\put(10,15){\line(1,0){5}}
\put(25,15){\vector(-1,0){10}}
\multiput(25,3)(0,2){6}{$\cdot$}
\end{picture}}
\put(30,30){
\begin{picture}(25,20)(0,0)
\multiput(0,3)(0,2){6}{$\cdot$}
\put(0,15){\line(1,0){5}}
\put(0,5){\vector(1,0){10}}
\put(15,15){\vector(-1,0){10}}
\put(15,5){\line(-1,0){5}}
\put(15,0){\framebox(10,20){}}
\end{picture}}
\put(68,25){
\begin{picture}(10,2)(0,1)
\put(-5,-1.5){$\bullet$}
\put(0,0){\line(1,0){10}}
\put(0,2){\line(1,0){10}}
\end{picture}}
\end{picture}
\nonumber \\ && +
\begin{picture}(65,50)(0,0)
\put(0,15){
\begin{picture}(30,20)(0,0)
\multiput(0,3)(0,2){6}{$\cdot$}
\put(0,5){\vector(1,0){10}}
\put(15,5){\line(-1,0){5}}
\put(0,15){\line(1,0){5}}
\put(15,15){\vector(-1,0){10}}
\end{picture}}
\put(15,0){
\begin{picture}(15,50)(0,0)
\put(15,35){\oval(20,20)[tl]}
\put(0,35){\oval(10,10)[br]}
\put(0,15){\oval(10,10)[tr]}
\end{picture} }
\put(40,0){
\begin{picture}(10,50)(0,0)
\put(0,25){\oval(20,40)[tr]}
\put(10,15){\line(0,1){10}}
\put(10,15){\vector(0,1){5}}
\put(10,30){\vector(0,1){5}}
\end{picture}}
\put(25,5){
\begin{picture}(20,40)(0,0)
\put(0,10){\oval(10,10)[b]}
\put(20,10){\oval(10,10)[b]}
\put(15,25){\line(0,-1){5}}
\put(5,25){\line(0,-1){5}}
\put(0,10){\framebox(20,10){}}
\put(5,25){\vector(0,1){10}}
\put(5,40){\line(0,-1){5}}
\put(15,25){\line(0,1){5}}
\put(15,40){\vector(0,-1){10}}
\multiput(3,38)(2,0){6}{$\cdot$}
\end{picture}}
\put(53,25){
\begin{picture}(10,2)(0,1)
\put(-5,-1.5){$\bullet$}
\put(0,0){\line(1,0){10}}
\put(0,2){\line(1,0){10}}
\end{picture}}
\end{picture}
+ 
\begin{picture}(105,50)(0,0)
\put(0,15){
\begin{picture}(30,20)(0,0)
\multiput(0,3)(0,2){6}{$\cdot$}
\put(0,5){\line(1,0){10}}
\put(15,5){\line(-1,0){5}}
\put(0,15){\line(1,0){5}}
\put(15,15){\vector(-1,0){10}}
\end{picture}}
\put(15,0){
\begin{picture}(15,50)(0,0)
\put(15,35){\oval(20,20)[tl]}
\put(15,15){\oval(20,20)[bl]}
\put(0,35){\oval(10,10)[br]}
\put(0,15){\oval(10,10)[tr]}
\put(15,25){\oval(10,20)[l]}
\end{picture} }
\put(70,0){
\begin{picture}(10,50)(0,0)
\put(15,25){\oval(10,20)[r]}
\put(0,25){\oval(10,40)[r]}
\put(15,15){\vector(-1,0){5}}
\put(15,15){\line(-1,0){15}}
\put(15,35){\line(-1,0){15}}
\end{picture}}
\put(30,0){
\begin{picture}(40,20)(0,0)
\multiput(40,3)(0,2){6}{$\cdot$}
\put(0,15){\line(1,0){5}}
\put(0,5){\vector(1,0){10}}
\put(15,15){\vector(-1,0){10}}
\put(15,5){\line(-1,0){5}}
\put(15,0){\framebox(10,20){}}
\put(25,5){\vector(1,0){10}}
\put(40,5){\line(-1,0){5}}
\put(25,15){\line(1,0){5}}
\put(40,15){\vector(-1,0){10}}
\end{picture}}
\put(30,30){
\begin{picture}(40,20)(0,0)
\multiput(0,3)(0,2){6}{$\cdot$}
\put(0,15){\line(1,0){5}}
\put(0,5){\vector(1,0){10}}
\put(15,15){\vector(-1,0){10}}
\put(15,5){\line(-1,0){5}}
\put(15,0){\framebox(10,20){}}
\put(25,5){\vector(1,0){10}}
\put(40,5){\line(-1,0){5}}
\put(25,15){\line(1,0){5}}
\put(40,15){\vector(-1,0){10}}
\end{picture}}
\put(93,25){
\begin{picture}(10,2)(0,1)
\put(-5,-1.5){$\bullet$}
\put(0,0){\line(1,0){10}}
\put(0,2){\line(1,0){10}}
\end{picture}}
\end{picture},
\end{eqnarray}
where the first line corresponds to the BS equation
at the BCS level.
The last line, which includes the meson exchange interactions, 
has a null effect in flavour vectors. It is relevant only for 
flavour singlets, and may include the axial anomaly.
%
%
\par
The quantitative results of this paper will be obtained with a particular 
chiral invariant strong potential, which is an extended version of the
Nambu and Jona-Lasinio potential \cite{Nambu}.
In Hadronic Physics the effective interaction should be simultaneously  
confining, in the Minkowsky space, local, and Lorentz invariant,  
but unfortunately no interaction which complies with all these  
constraints has yet been used to study chiral symmetry breaking. 
Our choice is to relax the last constraint, and use an instantaneous 
approximation which simplifies the energy dependence of the interaction. 
This approximation also has the advantage to allow a straightforward
application to condensed matter physics or low energy nuclear physics.
We use \cite{papfpi} a 2-body potential for Dirac quarks,
$(-{3 \over 4} {\vec{\lambda} \over 2}  
 \cdot {\vec{\lambda} \over 2})(\gamma_0 \cdot \gamma_0)
[K^3_0 (x-y)^2-U]\delta(t_x-t_y)$
where $U$ is an arbitrarily large constant and $\bf \lambda$
are the Gell-Mann matrices. 
The infinite $U$ reappears in the self energy, and 
for color singlet channels this cancels the 
infinitely attractive potential.
Any colored state will have a mass proportional to $U$ and will thus be
confined. 
The choice of an harmonic potential is not crucial, a linear or funnel 
potential have been also used, it simply has a suitable
Fourier transform  $-(2\pi)^3\Delta_k$.
Excellent results are obtained in a very broad range of hadronic
phenomena, with $K_0\simeq 250MeV$, excepting $f_{\pi}$ and
$\langle \bar \psi \psi \rangle$
which turn out \cite{papfpi} to be 
underestimated by factors of $4.5$ and $2^3$.
With an instantaneous interaction, we prefer to substitute the
Dirac fermions in terms of Weyl fermions, with propagators,
\begin{equation}
{\cal S}_{q}(w,\vec{k})={\cal S}_{\bar q}(w,\vec{k})=
{i \over w-E(k) +i\epsilon}
\end{equation}
which will be represented with an arrow pointing forward in the case of a  
quark, and an arrow pointing backward in the case of an antiquark moving
forward in the time direction.
This formalism is convenient to simplify the BS equation into the
Salpeter equation, in a form which is as close as possible to the
simpler Schr\"odinger equation.
When Weyl propagators are used, the vertices of the effective potential 
are redefined, they include the spinors $u$ and $v$.
The Dirac vertex $\gamma_0$ is now replaced by 
$u^{\dagger}u \ , \
u^{\dagger}v \ , \
v^{\dagger}u $ or $
v^{\dagger}v$ when the vertex is respectively connected to a quark,
a pair creation, a pair annihilation or an antiquark.
%
%
\par
Using the Weyl fermions, we expand the ladder in meson propagators
\begin{picture}(20,4)
\put(0,1){\line(1,0){20}}
\put(0,3){\line(1,0){20}}
\put(10,1.5){$_<$}
\end{picture}
, in Salpeter amplitudes $\phi^+ \ , \ \phi^-$ and in Bethe-Salpeter 
truncated amplitudes $\chi=\int V( u \phi^+v^\dagger+v \phi^-u^\dagger) $. 
We find that the self energy of the quark has a diagonal
component $\Sigma_d$ which contributes to the dynamical mass of the
quark,
\FL
\begin{eqnarray}
\label{4 diago}
&&\Sigma_d=
\begin{picture}(30,10)(0,0)
\put(0,0){\line(1,0){5}}
\put(25,0){\line(1,0){5}}
\put(5,0){\line(1,0){10}}
\put(25,0){\vector(-1,0){10}}
\put(5,0){$\cdot$}
\put(6,4.4){$\cdot$}
\put(8,7){$\cdot$}
\put(10.6,9){$\cdot$}
\put(15,10){$\cdot$}
\put(19.4,9){$\cdot$}
\put(22,7){$\cdot$}
\put(24,4.4){$\cdot$}
\put(25,0){$\cdot$}
\end{picture}  
+
\begin{picture}(30,10)(0,0)
\put(5,0){\line(4,-1){25}}
\put(25,0){\line(-4,-1){25}}
\put(5,0){\vector(1,0){10}}
\put(25,0){\line(-1,0){10}}
\put(5,0){$\cdot$}
\put(6,4.4){$\cdot$}
\put(8,7){$\cdot$}
\put(10.6,9){$\cdot$}
\put(15,10){$\cdot$}
\put(19.4,9){$\cdot$}
\put(22,7){$\cdot$}
\put(24,4.4){$\cdot$}
\put(25,0){$\cdot$}
\end{picture}
+ 
\begin{picture}(45,50)(0,0)
\put(10,05){
\begin{picture}(35,30)(0,0)
\put(4,9){\line(1,0){20}}
\put(4,11){\line(1,0){20}}
\put(20,2){$-E$}
\put(20,10){$_<$}
\put(24,30){\oval(22,42)[br]}
\put(24,28){\oval(18,34)[br]}
\end{picture}}
\put(0,0){
\begin{picture}(45,50)(0,0)
\put(0,10){\line(3,-1){15}}
\put(30,40){\vector(-1,-2){10}}
\put(0,20){\vector(3,1){15}}
\put(30,0){\vector(-3,1){15}}
\put(10,0){\line(1,2){15}}
\put(30,30){\line(-3,-1){15}}
\multiput(8,-2)(2,0){11}{$\cdot$}
\put(10,0){\line(-1,0){5}}
\put(30,0){\line(1,0){5}}
\end{picture}}
\put(0,5){
\begin{picture}(15,20)(0,0)
\put(0,0){\line(0,1){20}}
\put(0,0){\line(3,2){15}}
\put(0,20){\line(3,-2){15}}
\put(2,8){$\chi$}
\end{picture}}
\put(30,25){
\begin{picture}(15,20)(0,0)
\put(0,0){\line(0,1){20}}
\put(0,0){\line(3,2){15}}
\put(0,20){\line(3,-2){15}}
\put(2,8){$-$}
\end{picture}}
\end{picture} 
+
\begin{picture}(45,50)(0,0)
\put(10,5){
\begin{picture}(35,30)(0,0)
\put(4,9){\line(1,0){20}}
\put(4,11){\line(1,0){20}}
\put(20,2){$-E$}
\put(20,10){$_<$}
\put(24,30){\oval(22,42)[br]}
\put(24,28){\oval(18,34)[br]}
\end{picture}}
\put(0,0){
\begin{picture}(45,50)(0,0)
\put(0,10){\line(1,1){15}}
\put(30,40){\vector(-1,-1){15}}
\put(0,20){\vector(3,-2){25}}
\put(30,0){\line(-3,2){15}}
\put(10,0){\vector(2,3){15}}
\put(30,30){\line(-2,-3){15}}
\multiput(8,-2)(2,0){11}{$\cdot$}
\put(10,0){\line(5,-1){25}}
\put(30,0){\line(-5,-1){25}}
\end{picture}}
\put(0,5){
\begin{picture}(15,20)(0,0)
\put(0,0){\line(0,1){20}}
\put(0,0){\line(3,2){15}}
\put(0,20){\line(3,-2){15}}
\put(2,8){$\chi$}
\end{picture}}
\put(30,25){
\begin{picture}(15,20)(0,0)
\put(0,0){\line(0,1){20}}
\put(0,0){\line(3,2){15}}
\put(0,20){\line(3,-2){15}}
\put(2,8){$-$}
\end{picture}}
\end{picture} 
,
\end{eqnarray}
and the self energy of the antiquark is exactly the same.
In eq.(\ref{4 diago}) we only included the nonvanishing diagrams, of 
order $U$ or $1$.
The BCS quark propagators are proportional to $U^{-1}$.
The interactions without a pair creation or annihilation are proportional
to the infinite infrared constant $U$, while the remaining interactions
are finite.
It turns out that the diagrams with no creation or annihilation 
vertices are of order
$U$ while the diagrams with one finite interaction are finite.
The self energy also has antidiagonal components $\Sigma_a$ which
must vanish for the sake of vacuum stability.
These components would either create or annihilate scalar mesons 
in the vacuum.
The diagrams that contribute to the antidiagonal component of
the self energy in the mass gap equation are now,
\FL
\begin{eqnarray}
\label{4 antidiago}
\Sigma_a &=& 
\begin{picture}(30,10)(0,3)
\put(0,0){\line(1,0){5}}
\put(25,0){\line(-4,-1){25}}
\put(5,0){\line(1,0){10}}
\put(25,0){\vector(-1,0){10}}
\put(5,0){$\cdot$}
\put(6,4.4){$\cdot$}
\put(8,7){$\cdot$}
\put(10.6,9){$\cdot$}
\put(15,10){$\cdot$}
\put(19.4,9){$\cdot$}
\put(22,7){$\cdot$}
\put(24,4.4){$\cdot$}
\put(25,0){$\cdot$}
\end{picture}
+
\begin{picture}(30,10)(0,3)
\put(0,0){\line(1,0){5}}
\put(25,0){\line(-4,1){25}}
\put(5,0){\vector(1,0){10}}
\put(25,0){\line(-1,0){10}}
\put(5,0){$\cdot$}
\put(6,4.4){$\cdot$}
\put(8,7){$\cdot$}
\put(10.6,9){$\cdot$}
\put(15,10){$\cdot$}
\put(19.4,9){$\cdot$}
\put(22,7){$\cdot$}
\put(24,4.4){$\cdot$}
\put(25,0){$\cdot$}
\end{picture}
+
\begin{picture}(45,50)(0,0)
\put(10,5){
\begin{picture}(35,30)(0,0)
\put(4,9){\line(1,0){20}}
\put(4,11){\line(1,0){20}}
\put(20,2){$-E$}
\put(20,10){$_<$}
\put(24,30){\oval(22,42)[br]}
\put(24,28){\oval(18,34)[br]}
\end{picture}}
\put(0,0){
\begin{picture}(45,50)(0,0)
\put(0,10){\line(3,-1){15}}
\put(30,40){\vector(-1,-2){10}}
\put(0,20){\vector(3,1){15}}
\put(30,0){\vector(-3,1){15}}
\put(10,0){\line(1,2){15}}
\put(30,30){\line(-3,-1){15}}
\multiput(8,-2)(2,0){11}{$\cdot$}
\put(10,0){\line(-1,0){5}}
\put(30,0){\line(-5,-1){25}}
\end{picture}}
\put(0,5){
\begin{picture}(15,20)(0,0)
\put(0,0){\line(0,1){20}}
\put(0,0){\line(3,2){15}}
\put(0,20){\line(3,-2){15}}
\put(2,8){$\chi$}
\end{picture}}
\put(30,25){
\begin{picture}(15,20)(0,0)
\put(0,0){\line(0,1){20}}
\put(0,0){\line(3,2){15}}
\put(0,20){\line(3,-2){15}}
\put(2,8){$-$}
\end{picture}}
\end{picture}
+
\begin{picture}(45,50)(0,0)
\put(10,5){
\begin{picture}(35,30)(0,0)
\put(4,9){\line(1,0){20}}
\put(4,11){\line(1,0){20}}
\put(20,2){$-E$}
\put(20,10){$_<$}
\put(24,30){\oval(22,42)[br]}
\put(24,28){\oval(18,34)[br]}
\end{picture}}
\put(0,0){
\begin{picture}(45,50)(0,0)
\put(0,10){\line(1,1){15}}
\put(30,40){\vector(-1,-1){15}}
\put(0,20){\vector(3,-2){25}}
\put(30,0){\line(-3,2){15}}
\put(10,0){\vector(2,3){15}}
\put(30,30){\line(-2,-3){15}}
\multiput(8,-2)(2,0){11}{$\cdot$}
\put(10,0){\line(-1,0){5}}
\put(30,0){\line(-5,1){25}}
\end{picture}}
\put(0,5){
\begin{picture}(15,20)(0,0)
\put(0,0){\line(0,1){20}}
\put(0,0){\line(3,2){15}}
\put(0,20){\line(3,-2){15}}
\put(2,8){$\chi$}
\end{picture}}
\put(30,25){
\begin{picture}(15,20)(0,0)
\put(0,0){\line(0,1){20}}
\put(0,0){\line(3,2){15}}
\put(0,20){\line(3,-2){15}}
\put(2,8){$-$}
\end{picture}}
\end{picture} 
\nonumber \\ &&
+
\begin{picture}(65,45)(0,0)
\put(25,0){
\begin{picture}(35,30)(0,0)
\put(4,9){\line(1,0){20}}
\put(4,11){\line(1,0){20}}
\put(20,2){$-E$}
\put(20,10){$_<$}
\put(24,30){\oval(22,42)[br]}
\put(24,28){\oval(18,34)[br]}
\end{picture}}
\put(0,0){
\begin{picture}(45,50)(0,0)
\put(15,5){\line(-1,2){10}}
\put(45,35){\vector(-1,0){20}}
\put(15,15){\vector(3,1){15}}
\put(5,25){\vector(1,-2){5}}
\put(5,35){\line(1,0){20}}
\put(45,25){\line(-3,-1){15}}
\multiput(3,22)(0,2){6}{$\cdot$}
\put(5,25){\line(-1,0){5}}
\put(5,35){\line(-1,0){5}}
\end{picture}}
\put(15,0){
\begin{picture}(15,20)(0,0)
\put(0,0){\line(0,1){20}}
\put(0,0){\line(3,2){15}}
\put(0,20){\line(3,-2){15}}
\put(2,8){$\chi$}
\end{picture}}
\put(45,20){
\begin{picture}(15,20)(0,0)
\put(0,0){\line(0,1){20}}
\put(0,0){\line(3,2){15}}
\put(0,20){\line(3,-2){15}}
\put(2,8){$-$}
\end{picture}}
\end{picture}
+ 
\begin{picture}(65,45)(0,0)
\put(25,0){
\begin{picture}(35,30)(0,0)
\put(4,9){\line(1,0){20}}
\put(4,11){\line(1,0){20}}
\put(20,2){$-E$}
\put(20,10){$_<$}
\put(24,30){\oval(22,42)[br]}
\put(24,28){\oval(18,34)[br]}
\end{picture}}
\put(0,0){
\begin{picture}(45,50)(0,0)
\put(15,5){\line(1,1){15}}
\put(45,35){\vector(-1,-1){15}}
\put(15,15){\vector(-1,2){5}}
\put(5,25){\vector(1,0){20}}
\put(5,35){\line(1,-2){10}}
\put(45,25){\line(-1,0){20}}
\multiput(3,22)(0,2){6}{$\cdot$}
\put(5,25){\line(-1,0){5}}
\put(5,35){\line(-1,0){5}}
\end{picture}}
\put(15,0){
\begin{picture}(15,20)(0,0)
\put(0,0){\line(0,1){20}}
\put(0,0){\line(3,2){15}}
\put(0,20){\line(3,-2){15}}
\put(2,8){$\chi$}
\end{picture}}
\put(45,20){
\begin{picture}(15,20)(0,0)
\put(0,0){\line(0,1){20}}
\put(0,0){\line(3,2){15}}
\put(0,20){\line(3,-2){15}}
\put(2,8){$-$}
\end{picture}}
\end{picture}
+\dots
\end{eqnarray}
where the lines
\begin{picture}(20,0)(0,0) \put(0,0){\line(1,0){20}} \end{picture}
only mean the presence of a spinor $u, \ v, \ u^{\dagger}$ or
$v^{\dagger}$ and don't include the quark or anti-quark
Weyl propagator. The first pair of diagrams in eqs. (\ref{4 diago}),
(\ref{4 antidiago}) are BCS diagrams.
We will now concentrate on the new diagrams which depend on
the Coupled Channels. We will for instance consider the third diagram
of eq. (\ref{4 diago}). The integration of the quark propagators, which
are functions of the relative energies yields,
\begin{eqnarray}
&&\int {dw \over 2\pi} 
{i \over w -T_1 + i \epsilon}
{i \over E -w -T_2 + i \epsilon}
{i \over w -T_{1'} + i \epsilon}
\nonumber \\
&=&{i \over E -T_1 -T_2+ i \epsilon}
{i \over E -T_{1'} -T_2+ i \epsilon},
\end{eqnarray}
and we obtain a product of propagators $G_0$ of quark-antiquark pairs.
The right $G_0$ is in
fact part of the ladder and should be absorbed in the ket $\phi^-$.
We now integrate the remaining $G_0$, together with the propagator
of the ladder boundstate with energy $M_l(P)$, 
\begin{eqnarray}
&&\int {dE \over 2\pi}{i \over -E -M_l + i \epsilon}
{i \over E -T_1 -T_2+ i \epsilon}
\nonumber \\
&=&{i \over -M_l -T_1 -T_2+ i \epsilon}
\end{eqnarray}
which cancels the potential $-i V \equiv ( M_l -T_1 -T_2 ) / i$ that comes 
from the Salpeter equation 
(\ref{BCS BSE}).
After removing the poles in the propagators with this
integration in the loop energies,
we get the quark energy $T=u^\dagger \not k u -\Sigma_d$ where,
\FL
\begin{eqnarray}
\label{quark T}
\Sigma_d&=& u^\dagger \int {d^3k \over 2\pi^3}  V
[ { u u^\dagger - v v^\dagger \over 2 } +
u (\int {d^3P \over 2\pi^3}
\sum \phi^- {\phi^-}^{\dagger}) u^\dagger
\nonumber \\ &&
-
v (\int {d^3P \over 2\pi^3}
\sum \phi^- {\phi^-}^{\dagger}) v^\dagger
] u
\end{eqnarray}
where $P$ is the center of mass momentum of the coupled meson.
The mass gap equation $0=u^\dagger \not k v - \Sigma_a$ with,
\FL
\begin{eqnarray}
&\Sigma_a&= u^\dagger \int {d^3k \over 2\pi^3}  V
[ { u u^\dagger - v v^\dagger \over 2 } +
u (\int {d^3P \over 2\pi^3}
\sum \phi^- {\phi^-}^{\dagger}) u^\dagger 
\nonumber \\ &-&
v (\int {d^3P \over 2\pi^3}
\sum \phi^- {\phi^-}^{\dagger}) v^\dagger ]v+
\phi^- v ^\dagger \chi v + u ^\dagger \chi u {\phi^-}^{\dagger}.
\end{eqnarray}
Only the small component $\phi^-$ contributes to the result,
and this shows that the effect of the infinite
tower of coupled channels is finite.
In fact only the pion yields a sizeable contribution.
The contribution of the remaining
infinite series of mesons, which are excited states and are 
not Goldstone bosons, is negligible
because their $\phi^-\alpha M_l^{-1}$ decrease, and because
their average momentum increase.
%
%
\par
In eq. (\ref{quark T}), the infrared divergent kinetic energy, 
\begin{equation}
\label{masinf} 
T=U \left({1 \over 2} + \int{d^3P \over (2\pi)^3} \phi^- 
{\phi^-}^{\dagger} \right)+ finite \ terms .
\end{equation}
is now larger than in the BCS case, where it was simply $U/2$. 
We will now concentrate on the second line of eq.(\ref{CS3 kernel}),
and analyze the terms proportional to the infrared divergent $U$.
The diagrams with $\Sigma_s$ are clearly in that class, and the
last diagram also contributes in a subtle way. This diagram
can be divided in 4 different diagrams, which have a different
structure in the vertices of the interaction and in the quark
propagators. In Dirac notation, and skipping the arrows in the 
propagators for simplicity, we can sketch,
\FL
\begin{eqnarray}
\begin{picture}(30,30)(0,0)
\multiput(-3,8)(0,2){6}{$\cdot$}
\multiput(5,-2)(0,2){6}{$\cdot$}
\multiput(5,18)(0,2){6}{$\cdot$}
\put(12,-2){\framebox(5,14){}}
\put(12,18){\framebox(5,14){}}
\put(7,0){\line(1,0){5}}
\put(7,10){\line(1,0){5}}
\put(7,20){\line(1,0){5}}
\put(7,30){\line(1,0){5}}
\put(7,10){\oval(16,20)[bl]}
\put(7,20){\oval(16,20)[tl]}
\put(7,15){\oval(5,10)[l]}
\put(17,15){\oval(5,10)[r]}
\put(17,15){\oval(16,30)[r]}
\put(23,13){$\bullet$}
\put(25,14){\line(1,0){6}}
\put(25,16){\line(1,0){6}}
\end{picture}
=
\begin{picture}(20,30)(0,0)
\multiput(-3,8)(0,2){6}{$\cdot$}
\multiput(5,-2)(0,2){6}{$\cdot$}
\multiput(5,18)(0,2){6}{$\cdot$}
\put(7,10){\oval(16,20)[bl]}
\put(7,20){\oval(16,20)[tl]}
\put(7,15){\oval(5,10)[l]}
\put(7,15){\oval(5,10)[r]}
\put(7,15){\oval(16,30)[r]}
\put(12,13){$\bullet$}
\put(15,14){\line(1,0){6}}
\put(15,16){\line(1,0){6}}
\end{picture}
+
\begin{picture}(35,30)(0,0)
\multiput(-3,8)(0,2){6}{$\cdot$}
\multiput(5,-2)(0,2){6}{$\cdot$}
\multiput(5,18)(0,2){6}{$\cdot$}
\put(12,-2){\framebox(5,14){}}
\put(12,18){\framebox(5,14){}}
\put(7,0){\line(1,0){5}}
\put(7,10){\line(1,0){5}}
\put(7,20){\line(1,0){5}}
\put(7,30){\line(1,0){5}}
\multiput(20,-2)(0,2){6}{$\cdot$}
\multiput(20,18)(0,2){6}{$\cdot$}
\put(17,0){\line(1,0){5}}
\put(17,10){\line(1,0){5}}
\put(17,20){\line(1,0){5}}
\put(17,30){\line(1,0){5}}
\put(7,10){\oval(16,20)[bl]}
\put(7,20){\oval(16,20)[tl]}
\put(7,15){\oval(5,10)[l]}
\put(22,15){\oval(5,10)[r]}
\put(22,15){\oval(16,30)[r]}
\put(28,13){$\bullet$}
\put(30,14){\line(1,0){6}}
\put(30,16){\line(1,0){6}}
\end{picture}
+
\begin{picture}(35,30)(0,0)
\multiput(-3,8)(0,2){6}{$\cdot$}
\multiput(5,18)(0,2){6}{$\cdot$}
\put(12,18){\framebox(5,14){}}
\put(7,0){\line(1,0){15}}
\put(7,10){\line(1,0){15}}
\put(7,20){\line(1,0){5}}
\put(7,30){\line(1,0){5}}
\multiput(13,-2)(0,2){6}{$\cdot$}
\multiput(20,18)(0,2){6}{$\cdot$}
\put(17,20){\line(1,0){5}}
\put(17,30){\line(1,0){5}}
\put(7,10){\oval(16,20)[bl]}
\put(7,20){\oval(16,20)[tl]}
\put(7,15){\oval(5,10)[l]}
\put(22,15){\oval(5,10)[r]}
\put(22,15){\oval(16,30)[r]}
\put(28,13){$\bullet$}
\put(30,14){\line(1,0){6}}
\put(30,16){\line(1,0){6}}
\end{picture}
+
\begin{picture}(35,30)(0,0)
\multiput(-3,8)(0,2){6}{$\cdot$}
\multiput(5,-2)(0,2){6}{$\cdot$}
\put(12,-2){\framebox(5,14){}}
\put(7,0){\line(1,0){5}}
\put(7,10){\line(1,0){5}}
\put(7,20){\line(1,0){15}}
\put(7,30){\line(1,0){15}}
\multiput(20,-2)(0,2){6}{$\cdot$}
\multiput(12.5,18)(0,2){6}{$\cdot$}
\put(17,0){\line(1,0){5}}
\put(17,10){\line(1,0){5}}
\put(7,10){\oval(16,20)[bl]}
\put(7,20){\oval(16,20)[tl]}
\put(7,15){\oval(5,10)[l]}
\put(22,15){\oval(5,10)[r]}
\put(22,15){\oval(16,30)[r]}
\put(28,13){$\bullet$}
\put(30,14){\line(1,0){6}}
\put(30,16){\line(1,0){6}}
\end{picture}
,
\end{eqnarray}
and we find that the first diagram vanishes, the second is
finite and corresponds to the typical coupled channel
diagram, and the last two ones are of order $U$. Bringing
together the dominant diagrams, we find for the flavour vector channels,
with the Weyl notation,
\begin{eqnarray}
\begin{picture}(17,10)(0,0)
\put(2,0){
\begin{picture}(10,2)(0,-1)
\put(-5,-1.5){$\bullet$}
\put(0,0){\line(1,0){10}}
\put(0,2){\line(1,0){10}}
\end{picture}
}\end{picture}
\ &=& \ + \
\begin{picture}(30,10)(0,0)
\put(0,0){\vector(1,0){10}}
\put(10,0){\line(1,0){5}}
\put(15,5){\oval(10,10)[r]}
\put(15,10){\vector(-1,0){10}}
\put(5,10){\line(-1,0){5}}
\multiput(0,-2)(0,2){6}{$\cdot$}
\put(20,5){
\begin{picture}(10,2)(0,1)
\put(-5,-1.5){$\bullet$}
\put(0,0){\line(1,0){10}}
\put(0,2){\line(1,0){10}}
\end{picture}}
\end{picture}
\ +
\begin{picture}(55,40)(0,0)
\put(0,0){
\begin{picture}(35,30)(0,0)
\put(0,0){\line(1,0){35}}
\put(0,0){\vector(1,0){10}}
\put(35,5){\oval(10,10)[r]}
\put(35,10){\line(-1,1){20}}
\put(35,10){\vector(-1,1){5}}
\put(15,20){\line(1,0){20}}
\put(15,20){\vector(1,0){17}}
\put(35,30){\line(-1,-1){20}}
\put(15,10){\line(-1,0){15}}
\put(15,10){\vector(-1,0){10}}
\multiput(-2,-2)(0,2){6}{$\cdot$}
\put(38,3){$\bullet$}
\put(40,4){\line(1,0){10}}
\put(40,6){\line(1,0){10}}
\end{picture}}
\put(5,25){
\begin{picture}(40,15)(0,0)
\put(10,0){\oval(20,30)[tl]}
\put(10,2){\oval(16,22)[tl]}
\put(30,0){\oval(20,30)[tr]}
\put(30,2){\oval(16,22)[tr]}
\put(10,15){\line(1,0){20}}
\put(10,13){\line(1,0){20}}
\put(18,13){$_<$}
\put(15,6){$_{-E}$}
\end{picture}}
\put(5,15){
\begin{picture}(15,20)(0,0)
\put(0,10){\line(1,-1){10}}
\put(0,10){\line(1,1){10}}
\put(10,0){\line(0,1){20}}
\put(4,10){$_-$}
\end{picture}}
\put(35,15){
\begin{picture}(15,20)(0,0)
\put(0,0){\line(0,1){20}}
\put(0,0){\line(1,1){10}}
\put(0,20){\line(1,-1){10}}
\put(2,10){$_-$}
\end{picture}}
\end{picture}
+
\begin{picture}(55,40)(0,0)
\put(5,0){
\begin{picture}(35,30)(0,0)
\put(0,0){\line(1,0){35}}
\put(0,0){\vector(1,0){10}}
\put(35,5){\oval(10,10)[r]}
\put(35,10){\line(-1,1){20}}
\put(35,10){\vector(-1,1){5}}
\put(15,20){\line(1,0){20}}
\put(15,20){\vector(1,0){17}}
\put(35,30){\line(-1,-1){20}}
\put(15,10){\line(-1,0){15}}
\put(15,10){\vector(-1,0){10}}
\multiput(-2,-2)(0,2){6}{$\cdot$}
\multiput(18,-2)(0,2){11}{$\cdot$}
\put(5,30){\line(1,0){10}}
\put(5,20){\line(1,0){10}}
\put(5,20){\vector(1,0){10}}
\put(20,0){\vector(1,0){10}}
\put(38,3){$\bullet$}
\put(40,4){\line(1,0){10}}
\put(40,6){\line(1,0){10}}
\end{picture}}
\put(0,25){
\begin{picture}(50,15)(0,0)
\put(10,0){\oval(20,30)[tl]}
\put(10,2){\oval(16,22)[tl]}
\put(40,0){\oval(20,30)[tr]}
\put(40,2){\oval(16,22)[tr]}
\put(10,15){\line(1,0){30}}
\put(10,13){\line(1,0){30}}
\put(25,13){$_<$}
\put(22,6){$_{-E}$}
\end{picture}}
\put(0,15){
\begin{picture}(15,20)(0,0)
\put(0,10){\line(1,-1){10}}
\put(0,10){\line(1,1){10}}
\put(10,0){\line(0,1){20}}
\put(4,10){$_-$}
\end{picture}}
\put(40,15){
\begin{picture}(15,20)(0,0)
\put(0,0){\line(0,1){20}}
\put(0,0){\line(1,1){10}}
\put(0,20){\line(1,-1){10}}
\put(2,10){$_-$}
\end{picture}}
\end{picture}
\nonumber \\ && +
\begin{picture}(55,45)(0,0)
\put(0,0){
\begin{picture}(50,40)(0,0)
\put(0,40){\line(1,0){35}}
\put(15,40){\vector(-1,0){10}}
\put(35,35){\oval(10,10)[r]}
\put(35,30){\line(-1,-1){20}}
\put(25,20){\vector(1,1){7}}
\put(15,20){\line(1,0){20}}
\put(35,20){\vector(-1,0){7}}
\put(35,10){\line(-1,1){20}}
\put(15,30){\line(-1,0){15}}
\put(0,30){\vector(1,0){10}}
\multiput(-2,28)(0,2){6}{$\cdot$}
\put(38,33){$\bullet$}
\put(40,34){\line(1,0){10}}
\put(40,36){\line(1,0){10}}
\end{picture}}
\put(5,0){
\begin{picture}(50,15)(0,0)
\put(10,15){\oval(20,30)[bl]}
\put(10,13){\oval(16,22)[bl]}
\put(30,15){\oval(20,30)[br]}
\put(30,13){\oval(16,22)[br]}
\put(10,0){\line(1,0){20}}
\put(10,2){\line(1,0){20}}
\put(20,0){$_<$}
\put(17,6){$_{-E}$}
\end{picture}}
\put(5,5){
\begin{picture}(15,20)(0,0)
\put(0,10){\line(1,-1){10}}
\put(0,10){\line(1,1){10}}
\put(10,0){\line(0,1){20}}
\put(4,10){$_-$}
\end{picture}}
\put(35,5){
\begin{picture}(15,20)(0,0)
\put(0,0){\line(0,1){20}}
\put(0,0){\line(1,1){10}}
\put(0,20){\line(1,-1){10}}
\put(2,10){$_-$}
\end{picture}}
\end{picture}
+
\begin{picture}(55,45)(0,0)
\put(5,0){
\begin{picture}(50,40)(0,0)
\put(0,40){\line(1,0){35}}
\put(15,40){\vector(-1,0){10}}
\put(35,35){\oval(10,10)[r]}
\put(35,30){\line(-1,-1){20}}
\put(25,20){\vector(1,1){7}}
\put(15,20){\line(1,0){20}}
\put(35,20){\vector(-1,0){7}}
\put(35,10){\line(-1,1){20}}
\put(15,30){\line(-1,0){15}}
\put(0,30){\vector(1,0){10}}
\multiput(-2,28)(0,2){6}{$\cdot$}
\multiput(18,18)(0,2){11}{$\cdot$}
\put(5,10){\line(1,0){10}}
\put(5,20){\line(1,0){10}}
\put(20,20){\vector(-1,0){10}}
\put(35,40){\vector(-1,0){10}}
\put(38,33){$\bullet$}
\put(40,34){\line(1,0){10}}
\put(40,36){\line(1,0){10}}
\end{picture}}
\put(0,0){
\begin{picture}(50,15)(0,0)
\put(10,15){\oval(20,30)[bl]}
\put(10,13){\oval(16,22)[bl]}
\put(40,15){\oval(20,30)[br]}
\put(40,13){\oval(16,22)[br]}
\put(10,0){\line(1,0){30}}
\put(10,2){\line(1,0){30}}
\put(25,0){$_<$}
\put(22,6){$_{-E}$}
\end{picture}}
\put(0,5){
\begin{picture}(15,20)(0,0)
\put(0,10){\line(1,-1){10}}
\put(0,10){\line(1,1){10}}
\put(10,0){\line(0,1){20}}
\put(4,10){$_-$}
\end{picture}}
\put(40,5){
\begin{picture}(15,20)(0,0)
\put(0,0){\line(0,1){20}}
\put(0,0){\line(1,1){10}}
\put(0,20){\line(1,-1){10}}
\put(2,10){$_-$}
\end{picture}}
\end{picture}
\end{eqnarray}
and the dominant terms are,
\FL
\begin{eqnarray}
&(T_1+T_2)G_0\Gamma=U \left[ 1 +   \right. 
&2{T_1+T_2+T_3+T_4-M_l-U \over T_3 +T_4} 
\nonumber \\ 
&& \times \left. \int {d^3P \over (2\pi)^3} \phi^- {\phi^-}^{\dagger} 
\right] G_0\Gamma, 
\end{eqnarray} 
which is correct at order $U$, where  
$M_l = T_1+T_2+(-U)$. 
Thus we find a finite solution, with an exact cancelation of the terms   
proportional to $U$, although the infrared divergent
part of the kinetic energy is no longer $U / 2$. 
A cancelation also happens in flavour singlet channels.%
%
\par
We now study the finite effects of coupled channels.
Using the $\phi^-$ BCS amplitude of the $\pi$, we can estimate,
\begin{equation}
\label{2e}
\int{d^3P \over (2\pi)^3} \phi^- {\phi^-}^{\dagger} 
\simeq 2.5 N_f \ , \ \ N_f \simeq 3 \ ,
\end{equation}
which is large, due \cite{papfpi} to the small $f_{\pi}$, but finite. 
However if one goes beyond BCS, the ladder mesons pick up 
a large infinite mass according to Eq. (\ref{masinf}) 
, and this integral disappears because
$\phi^-\alpha M_l^{-1}\rightarrow 0$. 
If one then tries to solve the mass gap equation iteratively,
starting from the BCS solution, one finds 
a pair of accumulation points. 
The first is similar to the BCS solution, and the second
has a coupled channel contribution which renormalizes the
interaction, 
\FL
\begin{eqnarray}
\ \ T = {1 \over 2 }U \ ,& \ M_l=0U \ ,
& \ \int \phi^- {\phi^-}^{\dagger} = {\epsilon \over 2} 
\begin{picture}(40,1)(0,0) 
\put(35,-25){\oval(10,50)[tr]}
\put(40,-15){\vector(0,-1){10}}
\end{picture}
\nonumber \\
\begin{picture}(10,1)(0,0) 
\put(5,25){\oval(10,50)[bl]}
\put(0,15){\vector(0,1){10}}
\end{picture}
T = {1 + \epsilon \over 2 }U \ ,& \ M_l=\epsilon U \ ,
& \ \int {\phi^- {\phi^-}^{\dagger} 
\over (1 + 2 \epsilon)^2 }\alpha {1 \over M_l^2} = 0 
\end{eqnarray}
The stable intermediate solution is close to the BCS one,
it is obtained when
$\epsilon= (c / U)^{2/3}$ is a function of $U$,
\begin{equation}
M_l=\sqrt[3]{c^2U}, \ T = {U +\sqrt[3]{c^2U}\over 2 },
\ \int \phi^- {\phi^-}^{\dagger} = {c^2 \over 2 M_l^2}.
\end{equation}
In this case the $\phi^- \alpha U^{-{1 \over 3}}$ is suppressed, and the 
Salpeter equation for the ladder reduces essentially to a Schr\"odinger 
equation for the $\phi^+$ component. 
The coupled channel contribution to $\Sigma_a$ is vanishing,
of order $U^{-{2 \over 3}}$, and thus the mass gap equation is unchanged.
The remaining contributions from the coupled channels to the bound state 
equation are of order $M_l^{-1}$, and disappear
in the same way. 
As we predicted the mass of the bare (ladder) $\pi$ meson 
- which is now proportional to $\sqrt[3]{U}$ and infinite -
is exactly canceled by an opposite mass shift which 
is induced by the coupled channels, and the physical $\pi$ 
remains a Goldstone meson in the chiral limit.
\par
We developed an example
of a finite renormalizable field theory, and
there are good perspectives to apply this method to solid state physics.
The net result of including the (bare) coupled channels 
in eqs. (\ref{CS3 self},\ref{CS3 kernel}) is that no physical
deviation from the BCS case is found.
In order to get the finite contribution from coupled channels
we must either use a finite $U$ (nonconfining models), or
replace the bare (ladder) meson exchange in the mass gap equation 
(\ref{CS3 self}) by full meson exchange.  
It is possible to anticipate that in this model
the effect would be very large, in view of the result of eq (\ref{2e}) which 
would increase the quark condensate by a factor of $\simeq 2.5^3$.
In principle $f_{\pi}$ might also increase, thus improving the
physical results. 
\par
I am very grateful to Prof. J. Emilio Ribeiro 
for many long discussions and, in particular, for pointing the relevance
of the $\pi$ mass problem \cite{Emilio}. 
%

%
\end{document}